%% file: main.tex
\documentclass[manuscript]{acmart}

\newcommand{\re}[1]{{\color{black} #1}}

\usepackage{multirow}
\usepackage{subcaption}
\usepackage{caption}

\begin{document}

% \title[Design-Like Data Analysis through Parallel Prototyping and Rapid Iteration]{}

\title[Intelligent Canvas: Enabling Design-Like Exploratory Visual Data Analysis]{Intelligent Canvas: Enabling Design-Like Exploratory Visual Data Analysis with Generative AI through Rapid Prototyping, Iteration and Curation}

\author{Zijian Ding}
\affiliation{%
  \institution{University of Maryland, College Park}
  \country{USA}}

  \author{Joel Chan}
\affiliation{%
  \institution{University of Maryland, College Park}
  \country{USA}}
  
\renewcommand{\shortauthors}{Z. Ding et al.}

\begin{abstract}
Complex data analysis inherently seeks unexpected insights through exploratory \re{visual analysis} methods, transcending logical, step-by-step processing. However, \re{existing interfaces such as notebooks and dashboards have limitations in exploration and comparison for visual data analysis}. Addressing these limitations, we introduce a "design-like" intelligent canvas environment integrating generative AI into data analysis, offering rapid prototyping, iteration, and comparative visualization management. Our dual contributions include the integration of generative AI components into a canvas interface, and empirical findings from a user study (N=10) evaluating the effectiveness of the canvas interface.
\end{abstract}


\begin{teaserfigure}
  \includegraphics[width=\textwidth]{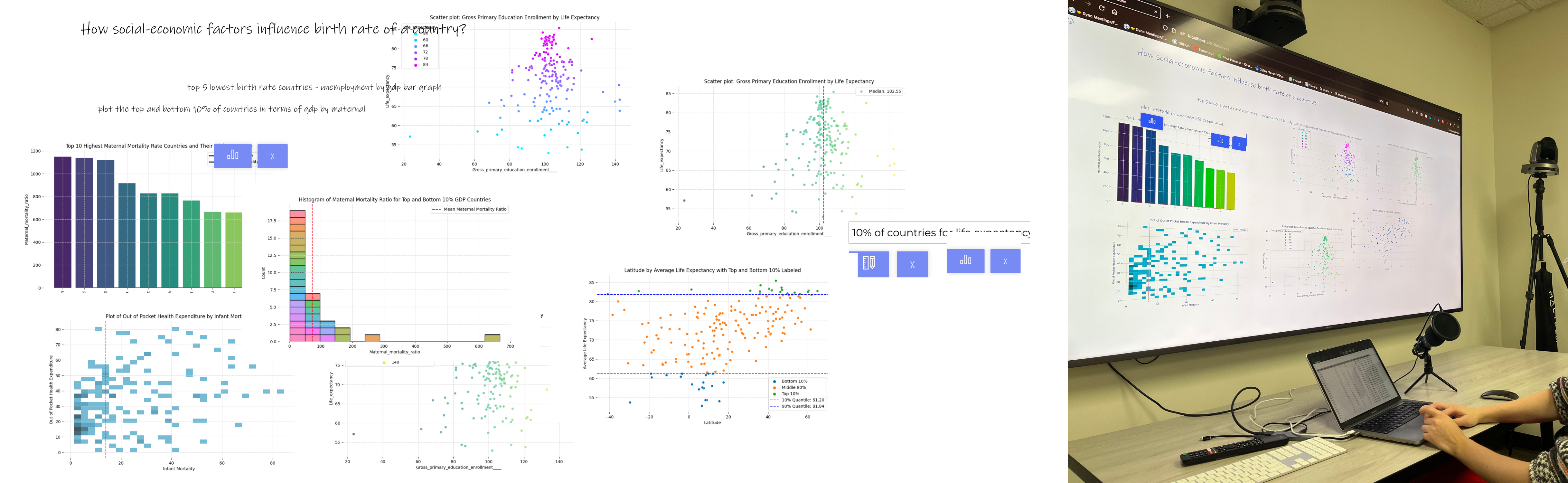}
  \caption{\re{Exploratory visual} data analysis results in design-like canvas environment: results (left) and study setting (right).
  % latitude by average life expectancy with top and bottom 10\% labeled
  }
  \Description{\re{Visual} data analysis results in design-like canvas environment.}
  \label{fig:P6}
\end{teaserfigure}

\maketitle

\def \RQO {\textbf{RQ1}: ?}

\def \RQT {\textbf{RQ2}: ?}

\input{sections/1-introduction}

\input{sections/2-relatedWork}

\input{sections/3-studyDesign}

\input{sections/4-results}

\input{sections/5-discussion_conclusion}

\input{sections/6-appendix}

\end{document}

%% file: sections/1-introduction.tex
\section{Introduction}
\label{sec:intro}

\re{

Complex data analysis presents a challenging task due to a multitude of factors such as an abundance of variables, the presence of competing hypotheses, and obscured causal relationships \cite{girolamiAnalysisComplexMultidimensional2006}. Take, for example, the question, ``what factors are contributing to the decline in video game sales", an economic downturn could result in a decrease in disposable income, potentially leading to a decline in video game sales as consumers gravitate towards free-to-play models. On the other hand, the same economic pressures might actually boost sales, with video games acting as a form of escapism. To effectively navigate this complexity, exploratory visual analysis is often favored over more linear, confirmatory approaches for its ability to represent data visually and interactively to enable intuitive insights \cite{keimChallengesVisualData2006}. Exploratory visual data analysis facilitates a more comprehensive understanding by enabling holistic thinking and sensemaking, thereby accommodating the intricate nature of complex data analysis \cite{battle2019characterizing}.

Existing platforms, while offering tools for visual data analysis, encounter limitations in effectively addressing its exploratory aspects \cite{liuBobaAuthoringVisualizing2021}. Interactive tools like Jupyter Notebooks, which combine code, data, and visualizations, fall short in enabling rapid iteration and comparative analysis of hypothesis test results due to their inherently linear and block-based structure \cite{chattopadhyayWhatWrongComputational2020b}. Similarly, data analysis dashboard platforms, such as Tableau\footnote{https://www.tableau.com/}, are designed to support parallel data display for comparative analysis but face challenges due to the complexity of their multi-functional interfaces. This intricacy may restrict the required space and flexibility for comparison, curation and sensemaking, as ``space to think" \cite{andrewsSpaceThinkLarge2010} in exploratory visual analysis, and potentially impede the speed and iterative nature essential for exploratory visual analysis \cite{wongsuphasawatVoyagerExploratoryAnalysis2016}.

% However, existing platforms exhibit limitations in addressing the exploratory dimensions of visual data analysis \cite{liuBobaAuthoringVisualizing2021}. Tools such as Jupyter Notebooks, representing an interactive paradigm to merging code, data, and visualizations, lack the capability for rapid iteration and comparison of hypothesis test results \cite{chattopadhyayWhatWrongComputational2020b}. Similarly, data analysis dashboard platforms such as Tableau\footnote{https://www.tableau.com/}, which facilitate parallel data display for comparative analysis, face challenges arising from the complexity of their interfaces for multiple different functions. However, the intricate nature of their interfaces, designed for various functions, can be constraining. This complexity may limit the necessary space and flexibility for comparison and curation in exploratory visual analysis. Additionally, it could affect the speed and iterative nature of processes required for exploratory visual analysis \cite{wongsuphasawatVoyagerExploratoryAnalysis2016}.

Recent progress in generative AI, notably large language models (LLMs) \cite{brownLanguageModelsAre2020}, offer new opportunities to overcome existing limitations in tools for exploratory visual data analysis. These models' capability in accurately converting high-level user intents into executable codes \cite{chenEvaluatingLargeLanguage2021} liberates data analysts from the intricacies of hands-on data visualization tasks, and facilitates rapid prototyping and iteration of visualizations, as exemplified by tools like OpenAI Code Interpreter\footnote{https://chat.openai.com/?model=gpt-4-code-interpreter} and LIDA\footnote{https://microsoft.github.io/lida/}. This advancement enables a shift in focus towards more abstract and high-level conceptual thinking, which is advantageous for exploratory visual data analysis. Additionally, the use of LLMs to interpret natural language inputs significantly reduces reliance on conventional graphical user interface (GUI), thus minimizing the space occupied by complex GUI elements such as sidebars and menus. This shift ushers in the development of a more expansive visual analysis workspace and flexible visualization provenance management interface \cite{callahanVisTrailsVisualizationMeets2006,silva2007provenance}, which is intrinsically beneficial for conducting activities like freeform curation \cite{kerneStrategiesFreeFormWeb2017, macneilFreeformTemplatesCombining2023, lupferPatternsFreeformCuration2016a}, sensemaking in data exploration \cite{texastechuniversityUsingVisualRepresentations2009}, and visual thinking \cite{arnheim2023visual}.

Those paradigm shifts in visual data analysis, increasingly akin to design methodologies such as rapid and iterative prototyping \cite{dowEfficacyPrototypingTime}, parallel prototyping \cite{dowParallelPrototypingLeads2010,dowEffectParallelPrototyping2009} and sensemaking \cite{russellCostStructureSensemaking1993} and free-form curation \cite{kerneStrategiesFreeFormWeb2017,macneilFreeformTemplatesCombining2023,lupferPatternsFreeformCuration2016a}, encourages us to ask the research question: \textbf{``How does a design-like canvas environment facilitate exploratory visual data analysis?"} To answer that question, we build a canvas-like environment for \re{visual} data analysis and study the . The canvas is a flexible and interactive environment for data analysts to perform rapid exploratory visual analysis and evaluate multiple potentially competing hypotheses. The proposed intelligent canvas encompasses three functionalities:

}

\begin{itemize}
    \item \textbf{Rapid Prototyping of Hypotheses:} We incorporate on-the-fly visualization features to hasten the transition from hypothesis generation to initial evaluation.
    \item \textbf{Rapid Iteration:} Our system architecture promotes iterative refinements and guided modifications, enabling the continuous evolution of hypotheses.
    \item \textbf{Visualization Management for Comparative Insight:} The canvas offers a freeform layout that allows users to juxtapose and curate hypotheses, thereby facilitating intuitive comparison and connection-making.
\end{itemize}

Our contributions are twofold, encompassing both technical and empirical aspects:

\begin{itemize}
\item \textbf{Technical Contribution}: We present the integration of generative AI components into \re{visual} data analysis, specifically through a canvas-like interface.
\item \textbf{Empirical Contribution}: We present empirical results from a user study, involving 10 participants who engaged with the generative AI-supported canvas for a 40-minute exploratory \re{visual} data analysis task. This was followed by a 20-minute semi-structured interview to gather in-depth insights into their experiences and interactions with the system.
\end{itemize}

%% file: sections/2-relatedWork.tex
\section{Related Work}

\re{

Our study builds upon existing research in interface design, expanding its scope by introducing an innovative canvas interface for visual data analysis tools. This development aims to tackle the challenges associated with exploration and comparison within visual data analysis, drawing upon design concepts.

Prior studies have addressed the limitations imposed by the linear layout of computational notebooks, which contrasts with unstructured and exploratory nature required for visual data analysis \cite{keryStoryNotebookExploratory2018,wang2022stickyland,weinmanForkItSupporting2021}. Besides, coding in notebooks is not sufficiently swift to accommodate the exploration of multiple threads in data analysis \cite{chenWHATSNEXTGuidanceenrichedExploratory}. To address those issues, prior research has extensively explored visualization recommendation systems, with a particular focus on dashboard interfaces, to enable swift and intuitive visual exploration of data.
Voyager provides faceted browsing of visualization recommendations based on statistical and perceptual measures \cite{wongsuphasawatVoyagerExploratoryAnalysis2016}.
Voyager 2 extends these capabilities by incorporating two partial view specifications: wildcards letting users specify multiple charts in parallel, and related views suggesting visualizations relevant to the currently specified chart \cite{wongsuphasawatVoyagerAugmentingVisual2017}.
Building upon dashboard-focused tools, Lyra, an interface to design customized visualizations for non-programmers, facilitates direct linkage of data to visual representations through drag-and-drop interactions \cite{satyanarayanLyraInteractiveVisualization2014}.
Lyra 2 extends Lyra to authoring interactive visualizations by demonstration \cite{zongLyraDesigningInteractive2020}. Chen et al. developed an interface to facilitate low-code exploration with insight-based guidance. This interface emphasizes recommendation panels and interactive tree visualizations that enable users to track analytic dependencies and navigate through their exploration history \cite{chenWHATSNEXTGuidanceenrichedExploratory}. 
 
In addition to dashboard tools tailored for exploratory visual data analysis, another notable trend is the development of multiverse analysis tools to implement all ``reasonable" analytic decisions concurrently. Boba, for instance, generates a multiplex of scripts representing all possible analysis paths from a singular codebase  \cite{liuBobaAuthoringVisualizing2021}. Sarma et al. \cite{sarmaMultiverseMultiplexingAlternative2021} developed an R tool that enables multiverse analyses through expressive syntax, integrates with notebook workflows, and increases efficiency by eliminating redundant computations. The proposition of debuggers tailored for multiverse analysis, along with tools like Causalvis that specialize in causal inference visualizations, suggest a concerted effort to enhance both the communicative clarity and iterative nature of data examination \cite{guUnderstandingSupportingDebugging2023, guoCausalvisVisualizationsCausal2023a}. Moreover, the development of B2 illustrates a blending of coding and visualization within computational notebooks, promoting an interactive, exploratory analysis experience \cite{wuB2BridgingCode2020}.

However, comparing data within the confines of a restrictive linear notebook or a cluttered and rigid dashboard interface filled with complex functionalities poses challenges. To mitigate the issue within the notebook framework, Wang et al. \cite{wang2022stickyland} introduced a "floating cells" solution, allowing segments of a notebook to be positioned freely across the screen. 
Alongside interface design advancements, the dimension of screen size in data analysis interfaces has garnered research interest. Studies like Knudsen's on large display benefits in data analysis \cite{knudsenExploratoryStudyHow2012} and Horak's work on combining smartwatches with large displays \cite{horakWhenDavidMeets2018} illuminate how larger screen sizes, as giving ``space to think" \cite{knudsenExploratoryStudyHow2012}, can help visual data analysis.

To address the challenges associated with exploration and comparison in exploratory visual data analysis, it is proposed to adopt methodologies from design practice, which inherently embody a highly exploratory process. This suggestion is underpinned by prior research that has adopted a design-centric approach, using a large whiteboard to simulate a substantial display area for data analysis, to examine how the utilization of ample display space can enhance data analysis activities \cite{knudsenExploratoryStudyHow2012}. 
Visual data analysis and design share multiple similarities, such as using visual elements to effectively convey information \cite{andrewSymbiosisDesignInformation}, adopting iterative processes for refinement \cite{nielsenIterativeUserinterfaceDesign1993,grammelHowInformationVisualization2010,wongsuphasawatVoyagerExploratoryAnalysis2016}, and emphasizing how the information is visually structured \cite{andrewSymbiosisDesignInformation,neumannFrameworkVisualInformation}. Prior research has integrated design concepts such as mock-ups -- rapid, low-fidelity prototypes -- into data visualizations, leading to a visualization mock-up toolkit \cite{vuillemotStructuringVisualizationMockUps2018}. However, a notable difference lies in execution ease; design often involves simpler sketching and drawing, whereas data visualization requires complex coding. Generative AI is changing that situation by translating user intents into executable codes for visualizations, thus aligning visual data analysis more closely with the intuitive processes of design, offering a new opportunity to develop a new interface based on the design concepts of parallel prototyping \cite{dowParallelPrototypingLeads2010, dowEffectParallelPrototyping2009} and free-form curation \cite{kerneStrategiesFreeFormWeb2017, macneilFreeformTemplatesCombining2023, lupferPatternsFreeformCuration2016a}. Our system expands upon those design concept to extend prior research in interface design for exploratory visual data analysis.

}

%% file: sections/3-studyDesign.tex
\section{System Design}

\re{

We proposed a set of design hypotheses and corresponding functionalities for the canvas interface:

}

\textbf{Rapid prototyping of hypotheses}: Our system streamlines the process of evaluation of \re{visual} data analysis \textit{hypotheses} by generating \textit{visualizations} with generative AI as hypotheses are typed in, referred to as rapid prototyping of hypotheses.

\textbf{Rapid iteration of visualization}: The architecture is built to facilitate iterative refinements to existing hypotheses. Guided modification makes it easy for users to evolve their hypotheses through iterative cycles.

\textbf{Visualization management for comparative insight}: To enable broader data understanding, our system incorporates a freeform canvas. This feature allows users to easily juxtapose and curate multiple hypotheses visually, facilitating intuitive comparison and interrelation. \re{To streamline this process, we have deliberately minimized interface elements that could obstruct that workflow, including restricting access to the code underlying the visualizations.}

\begin{figure}
\includegraphics[width=\textwidth]{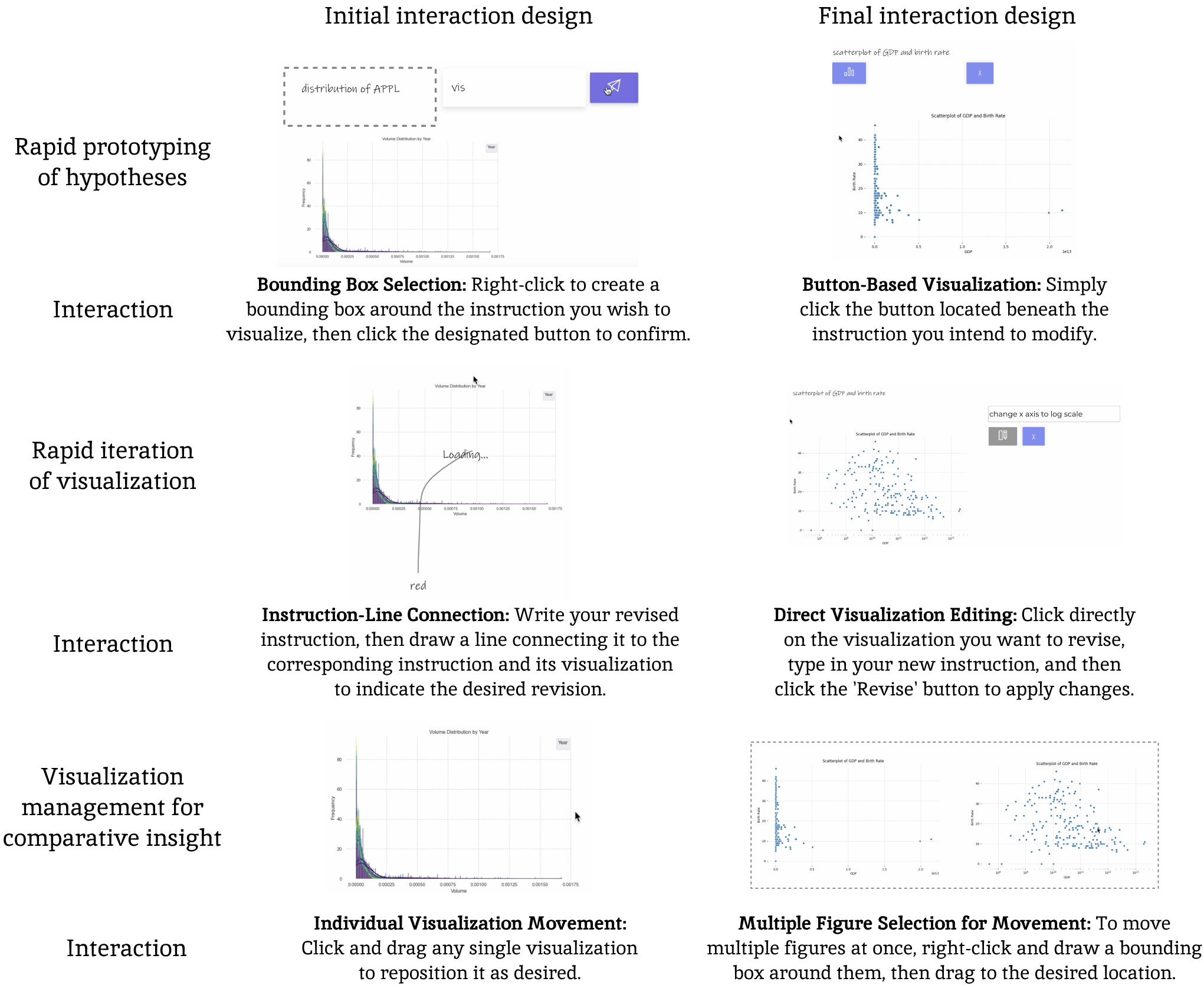}
  \caption{Iteration of interaction of the intelligent canvas corresponding to design hypotheses.}
  \Description{Iteration of functions of the intelligent canvas corresponding to three design hypotheses.}
  \label{fig:interaction}
\end{figure}

Guided by these hypotheses and functions, we underwent multiple iterations to refine the interactions, aiming to make them more direct and intuitive. This was done in accordance direct manipulation \cite{hutchinsDirectManipulationInterfaces} to minimize the user's effort in achieving their objectives. Figure \ref{fig:interaction} illustrates the evolution from initial to final interaction designs, focusing on three specific hypotheses. The initial design for rapid prototyping required users to input an instruction manually, followed by drawing a bounding box around it, a step which was somewhat extraneous. In pursuit of enhancing user efficiency and simplifying the interaction process, later iterations eliminated the need for creating a bounding box. This modification introduces a more efficient mechanism, allowing users to initiate the required function by simply clicking a button located immediately below the entered instruction. Likewise, the initial interaction for rapid iteration of visualizations required users to draw a line from the revision instruction to the center of the visualization, a process that was also found to be redundant. In response to this, the revised method eliminates the need to draw a line; users can simply click on the visualization they wish to modify and directly type the revising instruction in the prompt box.

The system utilized LIDA\footnote{https://microsoft.github.io/lida/} for its backend operations and was developed using React for its frontend interface.

\section{Methods}

% The "evaluation" stage of research through design.

\subsection{Participants Recruitment}

We recruited 10 participants aged between 24 and 37, of whom 9 were PhD students. Initially, the recruitment focused on individuals experienced in data analysis. However, early observations in the study revealed that participants with extensive data analysis experience tended to adhere to linear analysis process. In contrast, a participant with less experience exhibited more exploratory behavior. This observation prompted the recruitment of additional participants with less data analysis experience to enable a comparative analysis. As a result, we had 5 participants (P1-P5) with substantial data analysis experience (i.e., >=2 years with Python or R), and another 5 participants (P6-P10) with minimal to no experience (i.e., <2 years experience with Python or R).

\subsection{Dataset and Analysis Task}

The dataset utilized for the analysis task was the Global Country Information Dataset 2023\footnote{https://www.kaggle.com/datasets/nelgiriyewithana/countries-of-the-world-2023}, which encompasses a diverse range of variables, thereby presenting extensive opportunities for \re{visual} data analysis. This dataset comprises various columns, including GDP, GDP per capita, unemployment rate, life expectancy, maternal mortality ratio, minimum wage, out-of-pocket health expenditure, among others. We selected the open-ended analysis task, ``How do socio-economic factors, such as GDP per capita and minimum wage, influence the birth rate of a country?" to evaluate our design-like canvas interface in the context of complex \re{visual} data analysis tasks. The full analysis protocol is detailed in Appendix \ref{appendix:a}.

\subsection{Study Procedure}

Participants engaged in the study individually. Each session initiated with the signing of a consent form. The objective of the pre-survey was to capture demographic data and gauge the participant's prior experience with data analysis. After completing the pre-survey, participants were introduced to the intelligent canvas via a one-minute tutorial that covered all its functions. Throughout the study, they were also able to ask questions to a researcher for further clarification.

Subsequently, participants were tasked with exploring \re{visual} data analysis using the intelligent canvas on a large display, and also had the option to review the dataset on a separate screen of a laptop (see Figure \ref{fig:P6}). A time frame of 40 minutes was allotted to each participant for addressing the \re{visual} data analysis question, during which they utilized the intelligent canvas to execute the data analysis tasks.

During this phase, a "think-aloud" protocol was enforced, necessitating participants to vocalize their thought processes while interacting with the canvas. Both the screen interactions and participants' verbalizations were recorded for subsequent analysis.

After completing their tasks, participants engaged in a 20-minute semi-structured interview to offer further insights and feedback. The audio from these interviews, along with the think-aloud sessions, were transcribed in full using the OpenAI Whisper API\footnote{https://platform.openai.com/docs/guides/speech-to-text}. Subsequently, the first author carried out a thematic analysis \cite{braun2012thematic} of these transcripts, while simultaneously reviewing the participants' screen recordings. This involved open coding to identify emerging themes, such as parallel exploration and reducing exploratory costs. These themes were then cross-referenced with the three design hypotheses. For instance, themes like parallel exploration and reducing exploratory costs were linked to rapid prototyping of hypotheses.

\subsubsection{Semi-structured interview questions}

Below are the questions for the semi-structured interview:

\begin{itemize}
    \item Describe your experience with the intelligent canvas.
    \item How did the canvas's non-sequential layout affect your hypothesis exploration? Any specific insights gained?
    \item Evaluate the effectiveness of interaction methods (creating/changing/moving visualizations). Any limitations or new perspectives encountered? Suggestions for improvement?
    \item Compare the canvas with your usual analysis environment. Any specific features that were beneficial or disruptive in parallel exploration and interconnection discovery?
    \item Suggestions for improving spatial elements, visualization, or Generative AI functionalities for parallel hypothesis exploration?
    \item Would you adopt this tool for regular tasks or for what kinds of tasks? Why or why not? What analysis types or scenarios are most suited for it?
\end{itemize}

%% file: sections/4-results.tex
\section{Results}
\label{sec:results}

\subsection{Hypothesis 1: Rapid Prototyping of Hypotheses}

\subsubsection{Parallel Exploration by Participants with Extensive Data Analysis Experience}

Participants --- primarily with extensive
data analysis experience --- described
their perception of the value of the intelligent canvas for supporting parallel exploration (P2 with 10 years, P3 over 5 years, and P5 around 5 years of Python and R experience). For example, P2 contrasted their experiences with traditional programming environments to the multifaceted exploration enabled by the intelligent canvas:

\begin{quote}
    "With R or Python, we can work on just one plot at one time. I can now work on several, like put all of them there on the canvas, I can click, click which one I want to continue work. That is an advantage of this AI tool." (P2)
\end{quote}

Yet, this capacity of parallel prototyping was not without its challenges, as P5 noted. Despite the potential for increased productivity, the actual utilization of these features often fell short:

\begin{quote}
    "I wanted to do multiple things at once, and that would be kind of cool having a big whiteboard that you can plot, have multiple things going at once. Look at them side by side. But I guess I didn't really get to take advantage of that as much." (P5)
\end{quote}

Further elaborating on the exploration process, P3 discussed how the tool's features aligned with their approach to data exploration:

% Breadth-wise instead of depth-wise

\begin{quote}
    "It's like I could literally be doing all these different things at the same time, which I think is very good when it comes to exploration. Because you want, depending on what type of exploration you're doing, whether it's a depth or breadth type of exploration, I think this is very good for, like I said, the sequential, breadth-wise exploration of stuff. Depth-wise, maybe not. Breadth-wise, definitely, in terms of exploring all the things. But if I wanted to do a deeper analysis, like I said, probably not this one." (P3)
\end{quote}

\subsubsection{Reducing Exploratory Costs and Expanding Analysis Capabilities for Participants with Less Data Analysis Experience}

\begin{figure}
\includegraphics[width=\textwidth]{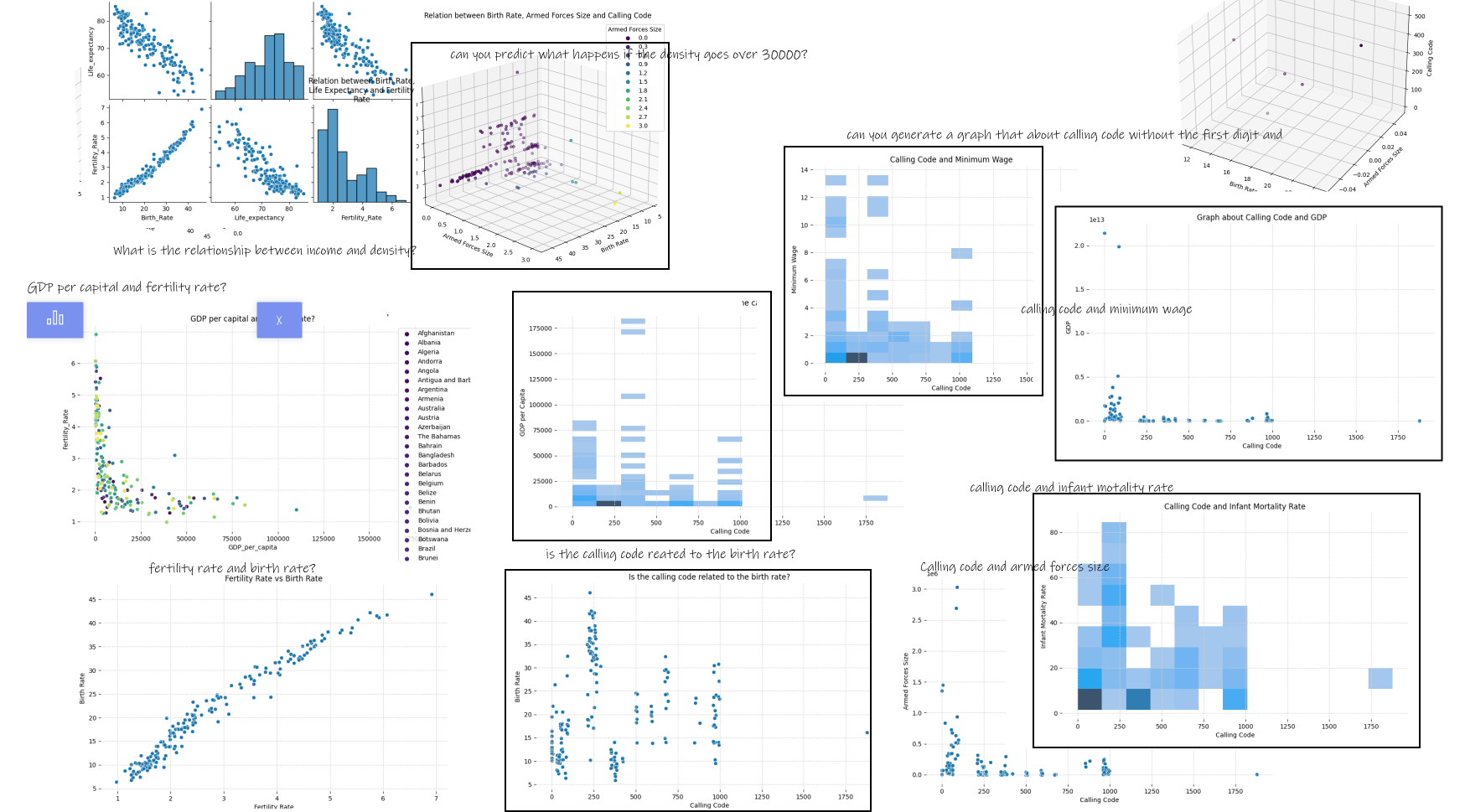}
  \caption{Rapid prototyping of P9 on the relationships between country calling codes and birth rates / GDP / GDP per capita / infant mortality rate / minimum wage / armed forces size.}
  \Description{Rapid prototyping of P9 on the relationships between country calling codes and birth rates / GDP / GDP per capita / infant mortality rate / minimum wage / armed forces size.}
  \label{fig:P9}
\end{figure}

Participants with limited experience in data analysis programming languages (P6 and P9, who had no experience with Python or R but used Nvivo), valued AI-generated data visualizations for their ability to lower exploratory costs and introduce functionalities absent in existing data analysis tools. P9 discussed the reduction in exploratory costs, enabling the testing of a wide range of hypotheses, including unconventional ones. This lowered cost of visualization fostered a greater ideational diversity:

\begin{quote}
"It lowers the cost of doing visualization to test out what visualization is better...Because you can see that I tested different hypotheses in the exploration stage. Even the most ridiculous hypothesis, like calling code with a birth rate. In fact, if I am doing that in the RStudio, I will not try that. Because each visualization requires you to do different codings. That limits my motivation of exploring some kind of sounds ridiculous ideas...It lowers the cost to explore ridiculous things, not just new things." (P9, Figure \ref{fig:P9})
\end{quote}

The capability of generative AI in creating visualizations that were previously challenging or even impossible with conventional tools was highlighted by P6. This participant conveyed a sense of liberation in being able to visualize data as desired:

\begin{quote}
    "it's definitely, in terms of the ability to see the data that I want to see, this blows that (another tool) out of the water by far. Because I can't even figure out how the hell to use the other one. And I have like three tutorials pulled up in a YouTube video right now, and it's not helping." (P6, Figure \ref{fig:P6})
\end{quote}

\subsection{Hypothesis 2: Rapid Iteration of Visualization}

\subsubsection{Enhancing Data Visualization Through Iteration}

The feature of prompt-driven rapid iteration proved instrumental for users aiming to enhance their data visualizations. The account from Participant P6 exemplified the stepwise refinement of complex visualizations through iterative modifications:

\begin{quote}
    "And once you get something that it gives you the picture, then you can start to play from there. I really like that, that there's the ability to modify it, because that's how we got the latitude by average life expectancy with a top and bottom 10\% label. If it hadn't given me the original one, and then let me say, okay, now flip it, and this, okay, now give me top and bottom 10, I would not have ever put in that as the initial query. And if I had put in that as the initial query, I don't know what would have happened. Because the first one that I put in that was that many chunks broke it. I'm guessing that it does better in step by step, rather than trying to parse it all at once." (P6, Figure \ref{fig:P6})
\end{quote}

Similarly, P10 highlighted the time-saving aspect of this iterative approach, appreciating the ability to directly manipulate and evolve existing charts:

\begin{quote}
    "You can pull out the graphs. You re-edit it on the previous one. It just saves you a lot of time by start over from scratch or whatever...I really like the function of you pulling the graph out of the original graph...You know, I really like the feature you can edit on the previous plot you had and regenerate a different plot." (P10)
\end{quote}

\subsubsection{Facilitating Comparative Analysis Through Iterative Visualization Retention}

The ability to retain and manipulate multiple iterations of visualizations on a single canvas offered a significant advantage in comparative analysis. P1 shared insights into how this feature provided a dynamic and integrated workspace, enhancing the user's ability to analyze and iterate on visual data:

\begin{figure}
  \includegraphics[width=\textwidth]{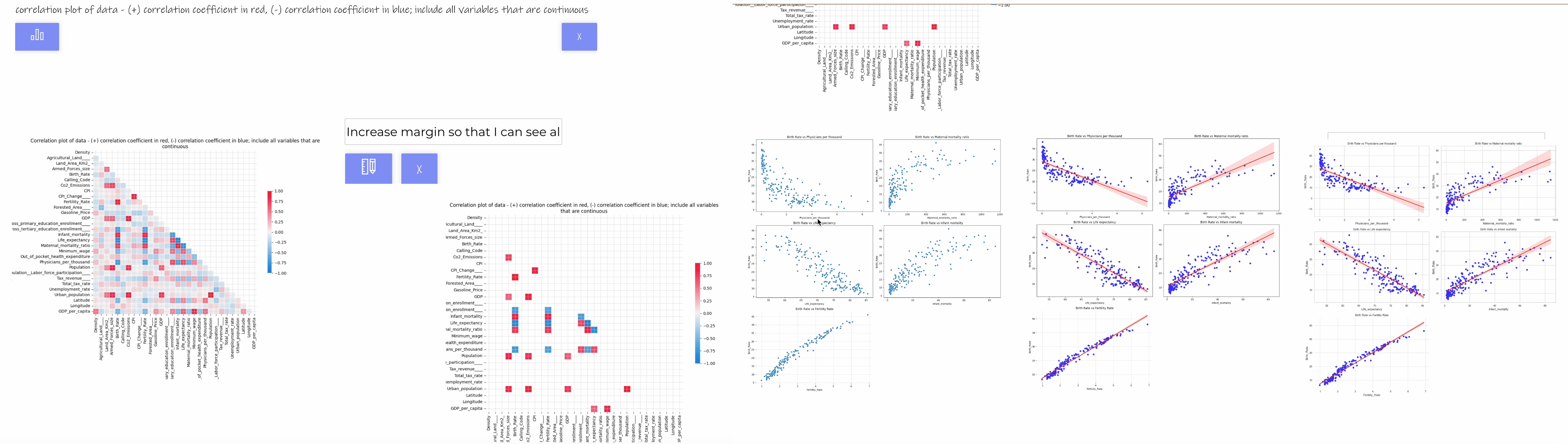}
  \caption{P1, generating an overview with correlation matrix (left) and then scatterplots for relationships with high correlations (right).}
  \label{fig:P1}
\end{figure}

\begin{quote}
    "It really is useful in the sense that I can see the steps. In Python, you draw one plot, and the plot is like...I have to open it to look at. But in the canvas, you can move it around. Then this one is like... You know, see the whole plots that I drew. And I can compare by not changing the pages or opening some new pages to see the other plot. I can just loop in this canvas and determine, this variable would be my key variable. Also, with this canvas, it's easier for you to use a figure as a reference. For example, use this figure as a reference and to create those subsequent plots. Maybe that's a good thing." (P1, Figure \ref{fig:P1})
\end{quote}

\subsection{Hypothesis 3: Visualization Management for Comparative Insight}

\subsubsection{Non-Sequential Canvas Workspace for Flexible Visualization Organization and Manipulation}

Participants discussed the benefits and challenges of using a non-linear, canvas-based approach for data visualization. They appreciated the flexibility and creative freedom this format provided for organizing and exploring data, yet also acknowledged the difficulties in managing a potentially chaotic layout. P10 highlighted the importance of a "free space" which allowed for a more dynamic and flexible approach to thinking and organizing thoughts or tasks.

\begin{quote}
    "Because human mindset is not working like sequentially, right? It's working here and there. At least my mindset is working like that. I need a free space like this." (P10)
\end{quote}

% \begin{quote}
%     "I'm not, I'm, okay, generally, I'm not a big fan of notebooks. Even when I'm doing my analysis, like for my data analysis, I generally do not use notebooks. I would rather do the scripts because, I don't know, it's just that sequential thing of, I have to, I'm not a big fan of it. But I guess that's what most of the tools are, kind of provide. And I kind of understand the premise, but I'm not a big fan of it. That's just the general summary. I prefer this. Because sometimes I don't want it to have to be sequential." (P3)
% \end{quote}

P10's remark emphasized the crucial role that visual appeal played in enriching the user experience and sparking creativity within the intelligent canvas, a sentiment echoed by P6, who appreciated the tool's automatic color defaults as fundamental visual elements.

\begin{quote}
    "Well, except the comparison, I really like, I think it just fits my aesthetics. It's beautiful, it's easy. You just click and create, right? You create in whatever way you like, rather than the tool list things for you." (P10)
\end{quote}

\begin{quote}
    "But I mean, overall, it's very fun, and I like that it automatically defaults to this series of colors." (P6, Figure \ref{fig:P6})
\end{quote}

P3 expressed enthusiasm about the concept of a blank canvas that allowed for spontaneous note-taking and questioning about a data set, which then led to the generation of various visualizations and charts.

\begin{quote}
    "I absolutely love this idea of having a blank canvas that you could kind of like just write down all the thoughts, the questions that you have about a data set and just see different visualizations and charts come up, right?" (P3)
\end{quote}

P6 compared the experience to using a Jamboard, emphasizing the ease of organizing information visually, like placing sticky notes. P6 valued the capability to visually categorize data with the flexibility and freedom that was often lacking in traditional spreadsheet applications such as Excel.

\begin{quote}
    "It made me treat it like Jamboard, like the sticky note approach, very much. Like every time I was doing something with it, I'm like, I'm just going to put this over here. And like I moved the two scatter plots, the gross primary, and then the one that I did to change the way the colors were to basically flip it X, Y based on how it was differentiating. Those two got to go stuck next to each other. Well, these two go together. It kind of like allows that sort of like mental categorization instead of having to rely on like, I've put this in this cell and now these cells are together. And I shouldn't sort these cells or I will ruin all of my data forever. And it was, it kind of felt like there was more room...But I think it's definitely, it like lends itself to a more like blank slate field and cells would have freaked me out if you had given me an Excel spreadsheet, this would have been kind of bad." (P6, Figure \ref{fig:P6})
\end{quote}

P10 valued the intelligent canvas for its simple, two-button interface that allowed for easy plot generation and fostered creative freedom:

\begin{quote}
    "And I really like this whole creative space. You can just click anywhere to generate a plot. And I like how there's only two buttons out there. It's just very easy to get your hands on generating plots." (P10)
\end{quote}

Moreover, participants recognized the freedom of manipulating and organizing visual elements to clearly trace and articulate their analytical thought processes. For instance, P3 found that this flexibility enhanced both their \re{visual} data analysis and presentation structuring:

\begin{quote}
    "And I like the fact that you can move things around, because then I can kind of like structure my analysis and even see the provenance of like my thought process as I'm going through, which I think is very important when you're doing analysis, because you can go back later and kind of like retrace the steps that you've taken. I definitely enjoyed working with this in general...And I think for me, even when I was thinking about this, it was kind of like, if you notice, I was kind of arranging the charts as I was going through. Because I was also thinking if I'm presenting this to somebody, I could kind of have certain charts in one section that are highlighting one insight, others in another section, right? And I could arrange all these things in however way I want to do it, giving me the freedom. If I want to do it sequentially, I can do it sequentially. If I don't want to do it sequentially, then I also want that freedom to not do it sequentially. I'm a very big fan of the canvas thing. I'm not even going to lie. When I was playing with it, I was like, huh, I think I want to do this in one of the tools I'm working on right now. I was like, I would love to see this whole canvas idea incorporated in there." (P3)
\end{quote}

P7 found that an expansive workspace enhanced their ability to group similar visualizations, which in turn assisted in identifying correlations among different relationships. They observed a resemblance between graphs pertaining to education and those related to physicians, indicating that the layout facilitated a comparative analysis.

\begin{quote}
    "The bigger space really helps me organize, you might have noticed this, how I was organizing like-looking graphs, then I can see the correlation. And when I noticed the one about the education, it looked very much like the one about the physicians." (P7)
\end{quote}

P8 specifically recognized the enhanced organizational capabilities afforded by a larger screen size, underscoring its positive impact on content management: "Having, definitely, the larger screen helps you with the organization."

Despite these advantages of a non-sequential workspace, some participants, like P4, P7, and P8, pointed out the challenges of managing a potentially chaotic layout, comparing it unfavorably to the structured environment of tools like Jupyter Notebook (P4), expressing difficulties in managing overlap and navigation (P7), and indicating a desire for more screen space to mitigate these issues (P8).

\begin{quote}
    "And it could be a very messy layout. It could be messier than Jupyter Notebook, like that? at least for Jupyter Notebook, you have a text box, right? Other than here, you just have one line of typing. And the Jupyter Notebook is like always 1, 2, 3, 4." (P4, Figure \ref{fig:P4})
\end{quote}

\begin{quote}
    "Well, I do like the space spreading it out, but then, you know, they get overlapped. Then I lose things. I think if I wanted to really research this topic, I would want like more screens." (P7)
\end{quote}

\begin{quote}
    "When you have many things arranged in a canvas, the whole panning and zooming thing can be hard, even with a large screen. I'm wondering how you can very smoothly move between those sorts of things." (P8)
\end{quote}

\subsubsection{Facilitating Intuitive Comparisons in Visualization Spaces}

Participants reflected on their experiences on how the intelligent canvas supports creative and intuitive comparison of visual data. Participants' quotes emphasized the advantage of being able to compare multiple analyses simultaneously, rather than in a linear, less interactive manner.

P3 articulated the benefits of an innovative workspace that allowed for concurrent comparisons:

\begin{quote}
    "Sometimes it could be like I'm doing two different things at the same time, right? Or I want to compare these two different threads of analysis side-by-side, not necessarily sequentially. This makes sense to me...I think especially for comparison, because like sometimes it could be, you could be checking certain variables or certain attributes in one particular site, right? And then I could have this other thing. I could also draw partitions. It doesn't even have to be side-by-side. I could do things like a grid of different analysis that I'm doing and kind of have them arranged in a way." (P3)
\end{quote}

P9 described the organization of visual outputs to facilitate comparative analysis:

\begin{quote}
    "But when it comes down, when I'm trying to explore the relationship between data, I started to mapping all this, or put all these images closely in some kinds of orderly, like here. If I move this away, you could see originally that I put the fertility rate versus negative birth rate next to the, just closely below the GDP per capita and the fertility rate. And also these four graphs, they could construct each other, contrast each other, I can know more about the details of the data...You can see this four figures down here. And I do place each close to each to another to see. Like, for example, the middle two figures, the calling code rate and the GDP per capita and the calling code with a burst rate. we could see there's a contrast between the 250 block of the calling code. This is a very adverse difference here. I do believe that the calling, this graph gives me the confidence of calling code is related to birth rate." (P9, Figure \ref{fig:P9})
\end{quote}

P10 offered insights into the superior readability offered by non-sequential visualization arrangements:

\begin{quote}
    "I think it just visually, I think it conveys better to me. For example, if I want to compare two graphs, two plots, I can just move the one next to the other. If you do sequential way, it's always underneath the other one. It's not readable. That is really just doing a better job." (P10)
\end{quote}

\begin{figure}
  \includegraphics[width=\textwidth]{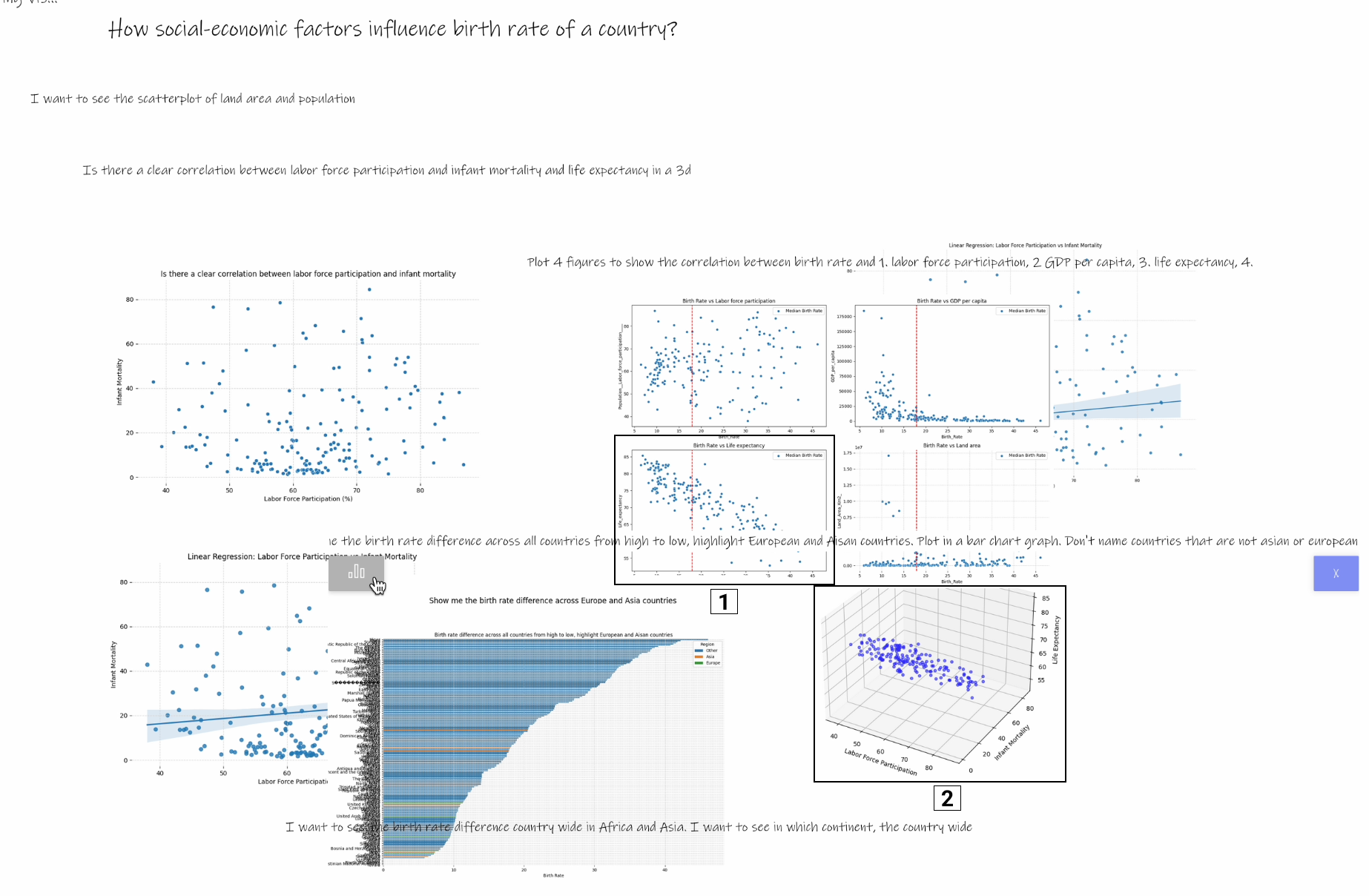}
  \caption{P4, comparing (1) birth rate and life expectancy and (2) 3D plot of labor force participation, infant mortality and life expectancy.}
  \label{fig:P4}
\end{figure}

P4 pointed out the inherent difficulties in juxtaposing images for comparison within a Jupyter Notebook, where visualizations are often displayed in a linear fashion. The participant expressed a preference for the AI visualization canvas, which simplified this process by allowing figures to be placed side by side for immediate visual comparison:

\begin{quote}
    "Well, for Jupyter Notebook, it's hard to put the figures together to compare. I need to compare this figure with the one I plotted downwards, right? For these kind of visualizations, I would prefer Canvas." (P4, Figure \ref{fig:P4})
\end{quote}

P5 provided insight into the usability challenges with Jupyter Notebook when attempting to arrange and compare visual elements. The intelligent canvas, by contrast, appeared to offer a more intuitive interface that facilitated the direct manipulation of visual elements, leading to a more convenient and satisfying user experience. This ease of rearrangement on the canvas was contrasted with the more rigid and less user-friendly cell manipulation in Jupyter Notebook.

\begin{quote}
    "I think there were some things I wanted to do, like to drag things and to compare, like even if I wasn't doing it, even if I was just looking at two different scatter plots, being able to rearrange and look at them seems like it might be a little more convenient than in a Jupyter notebook. I feel like maybe that's just because Jupyter notebooks are kind of annoying, like manipulating where the cells are and things." (P5)
\end{quote}

However, P6 valued the flexibility in organizing visualizations but advised against directly comparing distinct data sets like unemployment and GDP, to avoid misinterpretation due to complex underlying factors.

\begin{quote}
    "I like that I can move them around. I found that I didn't necessarily want to compare them, though. Because it's almost apples and oranges when you're talking about demography. It's different pieces of the picture, but it's not necessarily things that you would ever want to directly compare, because you might accidentally inflate things. If I'm looking at, if I have a chart that's, I don't know, unemployment, and I also have a chart that's GDP, those two are related. But I don't necessarily want to put them next to each other and then say, these go together this way, because there are many other qualitative sociological factors in there that that's almost a dangerous statement to take if you don't have the other data involved." (P6, Figure \ref{fig:P6})
\end{quote}

% \subsection{User behaviors}

% \subsubsection{Interactions with intelligent canvas}

% \subsubsection{Comparison with previous data analysis behaviors}

% \subsubsection{Comparison between experts and novices}

% \subsection{Features users appreciated, areas requiring refinement, and prospective new functionalities}

% \subsubsection{Features users appreciated}

% \subsubsection{Areas requiring refinement}

% \subsubsection{Prospective new functionalities}

%% file: sections/5-discussion_conclusion.tex
\section{Discussion}

\subsection{Summary and Interpretation of Results}

This research seeks to understand participant engagement with three functionalities in a \re{visual} data analysis task using the intelligent canvas: rapid prototyping of hypotheses, rapid iteration of visualization, and visualization management for comparative insight.

For rapid prototyping of hypotheses, results suggest that the intelligent canvas potentially enhances data visualization by enabling parallel exploration, particularly benefiting those with extensive data analysis experience. This feature allows for the simultaneous management of multiple plots, which may increase efficiency. For participants with less data analysis experience, the application of AI in this context appears to reduce exploratory costs, potentially expanding the range of hypotheses that can be tested, including unconventional ones. The capabilities of generative AI in this environment seem to surpass those of traditional tools, offering more sophisticated visualization possibilities.

Secondly, for rapid iteration of visualization, the study indicates that prompt-driven iterative processes could improve the development of complex visualizations. This approach allows users the flexibility for gradual refinements, potentially leading to more polished visual outputs. The efficiency of this iterative method is noteworthy, as it enables users to adapt and evolve their charts more rapidly. Another aspect of this process is the ability to maintain and adjust multiple visualization iterations on a single canvas, which may facilitate comparative analysis.

Regarding visualization management for comparative insight, the results emphasize participants' preference for a dynamic and flexible workspace, like the ability to place analyses side-by-side for comparison, a feature less common in linear tools such as Jupyter Notebook. Such spatial organization may assist participants in identifying patterns and relationships within the visualized data, though caution is advised in directly comparing different data entries to avoid erroneous interpretations. The aesthetic appeal and user-friendly interface of the canvas are observed to potentially enhance user experience and stimulate creativity. \re{Nevertheless, the absence of a structure to manage visualizations within the open canvas underscores the complexity of navigating intricate and potentially chaotic visual layouts, particularly when the canvas is saturated with an abundance of elements.}

\re{

\subsection{Experience-Based Customization: More Control for the Experienced, Extra Guidance for Beginners}

In addition to findings corresponding to the three primary hypotheses, the study also revealed variations in user preferences for the tool, influenced by their differing levels of experience in data analysis. Specifically, more experienced participants expressed a desire for greater control over the visualization generation process. This included access to the underlying code for various reasons: to understand how the visualizations were created, to save the code for future use, and to directly modify the code for iterative improvements. Furthermore, participants, particularly those with less experience in data analysis, sought increased support and guidance. They expressed a need for initial scaffolding to address the cold start problem, such as clear explanations of the model's capabilities, starting options, examples of effective prompts, and suggestions for prompt refinement and completions.

}

\subsection{Generalization: Developing the Next Generation of Graphical User Interface for Generative AI}

This study, analyzing participant use of a novel, flexible interface for \re{visual} data analysis, provides insights for designing graphical user interfaces (GUIs) in Generative AI environments. Moving away from the linear, command-line-like structure of platforms like Jupyter Notebook, the intelligent canvas showcases a shift towards more dynamic GUI approaches in \re{visual} data analysis. Essential to future GUI development are two key functions: rapid and iterative prototyping, and spatial content management. The former capitalizes on AI's capability for quick content creation, while the latter utilizes both the volume of AI-generated content and the GUI's spatial features. These functions hold significance not only in \re{visual} data analysis but also in other domains of generative AI, such as text-to-image generation \cite{bradePromptifyTexttoImageGeneration2023}. Insights from this study are poised to influence the broader design of generative AI systems and tools, such as harnessing AI's capabilities to enhance user creativity by reducing exploration costs, and addressing challenges like managing complex layouts. 
% Challenge for Generative AI support tools for experienced users: fixation to previous tools

\subsection{Limitations and Future Work}

This study's limitations are identified in three areas: 1) existing constraints of generative AI algorithms as of November 2023, such as latency, the inability to process multiple visualization requests simultaneously (hindering parallel prototyping experiences), and sporadic generation failures; 2) limitations in the current minimum viable prototype, especially its handling of \re{visual} data analysis code snippets for visualization generation, which remain inaccessible for direct user editing, and the lack of progress feedback during content generation; 3) the open-ended and exploratory nature of the study, limiting the ability to measure performance differences between traditional \re{visual} data analysis tools and the intelligent canvas. To address these challenges, future research could focus on the implementation of advanced generative AI algorithms and the improvement of user control and feedback mechanisms. This approach would enable users to monitor the progress of visualization generation and actively manipulate the process by providing access to and the ability to edit the underlying code. Additionally, adopting a more controlled study design, like a between-subjects comparison, to more effectively evaluate the effectiveness of functionalities in Generative AI-based tools.

\section{Conclusion}

This research demonstrates the potential of a "design-like" canvas system integrating Generative AI for complex \re{visual} data analysis, addressing the need for advanced exploratory methods. The system's capabilities in rapid prototyping, iterative visualization, and comparative insight management align with the challenges identified in traditional \re{visual} data analysis platforms. Findings from the user study indicate an enhancement in efficiency and a facilitation of a more effective \re{visual} analysis process for users with varying levels of experience. However, the research also identifies areas for improvement, particularly in layout management and screen space optimization. Overall, these results contribute to the evolving field of \re{visual} data analysis, suggesting a valuable direction for ongoing research of building interface for \re{Generative AI}.

% anonymity check
% \begin{acks}

% \end{acks}

\bibliographystyle{ACM-Reference-Format}
\bibliography{sample-base}

% \appendix

% \section{Appendix A}

% \section{Appendix B}

%% file: sections/6-appendix.tex
\appendix
\newpage
\textbf{APPENDIX}

\section{Data analysis protocol}
\label{appendix:a}

The Global Country Information Dataset 2023 offers a rich array of variables that provide substantial avenues for data analysis and visualization.

Here are the columns of the dataset: Country, Density, Agricultural Land(\%), Land Area(Km2), Armed Forces size, Birth Rate, Co2-Emissions, Fertility Rate, Forested Area(\%), Gasoline Price, GDP, Gross primary education enrollment(\%), Gross tertiary education enrollment(\%), Infant mortality, Life expectancy, Maternal mortality ratio, Minimum wage, Out of pocket health expenditure, Physicians per thousand, Population, Population: Labor force participation(\%), Tax revenue(\%), Total tax rate, Unemployment rate, Urban\_population, GDP per capita.

\textbf{Question to be answered by data visualization}

\textit{How social-economic factors e.g. GDP per capita, minimum wage influence the birth rate of a country?}

The interplay between birth rates and socio-economic indicators like GDP per capita and minimum wage is complex and nuanced. Higher GDP doesn't always mean higher birth rates, as factors like education and living costs come into play. Meanwhile, minimum wage affects financial security, swaying family size decisions. By dissecting the data in Global Country Information Dataset 2023, we can unravel intricate relationships and gain fresh insights into the forces shaping global demographic trends. So, analysts, ready for the challenge?

%% file: main.bbl

\begin{thebibliography}{41}

%%% the Web address.

\ifx \showCODEN    \undefined \def \showCODEN     #1{\unskip}     \fi
\ifx \showDOI      \undefined \def \showDOI       #1{#1}\fi
\ifx \showISBNx    \undefined \def \showISBNx     #1{\unskip}     \fi
\ifx \showISBNxiii \undefined \def \showISBNxiii  #1{\unskip}     \fi
\ifx \showISSN     \undefined \def \showISSN      #1{\unskip}     \fi
\ifx \showLCCN     \undefined \def \showLCCN      #1{\unskip}     \fi
\ifx \shownote     \undefined \def \shownote      #1{#1}          \fi
\ifx \showarticletitle \undefined \def \showarticletitle #1{#1}   \fi
\ifx \showURL      \undefined \def \showURL       {\relax}        \fi
% invisible to TeX
\providecommand\bibfield[2]{#2}
\providecommand\bibinfo[2]{#2}
\providecommand\natexlab[1]{#1}
\providecommand\showeprint[2][]{arXiv:#2}

\bibitem[\protect\citeauthoryear{Andrew}{Andrew}{[n.d.]}]%
        {andrewSymbiosisDesignInformation}
\bibfield{author}{\bibinfo{person}{VANDE~MOERE Andrew}.} \bibinfo{year}{[n.d.]}\natexlab{}.
\newblock \showarticletitle{The {Symbiosis} between {Design} and {Information} {Visualization}}.
\newblock  (\bibinfo{year}{[n.\,d.]}).
\newblock


\bibitem[\protect\citeauthoryear{Andrews, Endert, and North}{Andrews et~al\mbox{.}}{2010}]%
        {andrewsSpaceThinkLarge2010}
\bibfield{author}{\bibinfo{person}{Christopher Andrews}, \bibinfo{person}{Alex Endert}, {and} \bibinfo{person}{Chris North}.} \bibinfo{year}{2010}\natexlab{}.
\newblock \showarticletitle{Space to {Think}: {Large} {High}-resolution {Displays} for {Sensemaking}}. In \bibinfo{booktitle}{\emph{Proceedings of the {SIGCHI} {Conference} on {Human} {Factors} in {Computing} {Systems}}} \emph{(\bibinfo{series}{{CHI} '10})}. \bibinfo{publisher}{ACM}, \bibinfo{address}{New York, NY, USA}, \bibinfo{pages}{55--64}.
\newblock
\showISBNx{978-1-60558-929-9}
\urldef\tempurl%
\url{https://doi.org/10.1145/1753326.1753336}
\showDOI{\tempurl}


\bibitem[\protect\citeauthoryear{Arnheim}{Arnheim}{2023}]%
        {arnheim2023visual}
\bibfield{author}{\bibinfo{person}{Rudolf Arnheim}.} \bibinfo{year}{2023}\natexlab{}.
\newblock \bibinfo{booktitle}{\emph{Visual thinking}}.
\newblock \bibinfo{publisher}{Univ of California Press}.
\newblock


\bibitem[\protect\citeauthoryear{Battle and Heer}{Battle and Heer}{2019}]%
        {battle2019characterizing}
\bibfield{author}{\bibinfo{person}{Leilani Battle} {and} \bibinfo{person}{Jeffrey Heer}.} \bibinfo{year}{2019}\natexlab{}.
\newblock \showarticletitle{Characterizing {Exploratory} {Visual} {Analysis}: {A} {Literature} {Review} and {Evaluation} of {Analytic} {Provenance} in {Tableau}}.
\newblock \bibinfo{journal}{\emph{Computer Graphics Forum}} \bibinfo{volume}{38}, \bibinfo{number}{3} (\bibinfo{date}{June} \bibinfo{year}{2019}), \bibinfo{pages}{145--159}.
\newblock
\showISSN{0167-7055, 1467-8659}
\urldef\tempurl%
\url{https://doi.org/10.1111/cgf.13678}
\showDOI{\tempurl}


\bibitem[\protect\citeauthoryear{Brade, Wang, Sousa, Oore, and Grossman}{Brade et~al\mbox{.}}{2023}]%
        {bradePromptifyTexttoImageGeneration2023}
\bibfield{author}{\bibinfo{person}{Stephen Brade}, \bibinfo{person}{Bryan Wang}, \bibinfo{person}{Mauricio Sousa}, \bibinfo{person}{Sageev Oore}, {and} \bibinfo{person}{Tovi Grossman}.} \bibinfo{year}{2023}\natexlab{}.
\newblock \showarticletitle{Promptify: {Text}-to-{Image} {Generation} through {Interactive} {Prompt} {Exploration} with {Large} {Language} {Models}}. In \bibinfo{booktitle}{\emph{Proceedings of the 36th {Annual} {ACM} {Symposium} on {User} {Interface} {Software} and {Technology}}}. \bibinfo{publisher}{ACM}, \bibinfo{address}{San Francisco CA USA}, \bibinfo{pages}{1--14}.
\newblock
\showISBNx{9798400701320}
\urldef\tempurl%
\url{https://doi.org/10.1145/3586183.3606725}
\showDOI{\tempurl}


\bibitem[\protect\citeauthoryear{Braun and Clarke}{Braun and Clarke}{2012}]%
        {braun2012thematic}
\bibfield{author}{\bibinfo{person}{Virginia Braun} {and} \bibinfo{person}{Victoria Clarke}.} \bibinfo{year}{2012}\natexlab{}.
\newblock \bibinfo{booktitle}{\emph{Thematic analysis.}}
\newblock \bibinfo{publisher}{American Psychological Association}.
\newblock


\bibitem[\protect\citeauthoryear{Brown, Mann, Ryder, Subbiah, Kaplan, Dhariwal, Neelakantan, Shyam, Sastry, Askell, Agarwal, Herbert-Voss, Krueger, Henighan, Child, Ramesh, Ziegler, Wu, Winter, Hesse, Chen, Sigler, Litwin, Gray, Chess, Clark, Berner, McCandlish, Radford, Sutskever, and Amodei}{Brown et~al\mbox{.}}{2020}]%
        {brownLanguageModelsAre2020}
\bibfield{author}{\bibinfo{person}{Tom~B. Brown}, \bibinfo{person}{Benjamin Mann}, \bibinfo{person}{Nick Ryder}, \bibinfo{person}{Melanie Subbiah}, \bibinfo{person}{Jared Kaplan}, \bibinfo{person}{Prafulla Dhariwal}, \bibinfo{person}{Arvind Neelakantan}, \bibinfo{person}{Pranav Shyam}, \bibinfo{person}{Girish Sastry}, \bibinfo{person}{Amanda Askell}, \bibinfo{person}{Sandhini Agarwal}, \bibinfo{person}{Ariel Herbert-Voss}, \bibinfo{person}{Gretchen Krueger}, \bibinfo{person}{Tom Henighan}, \bibinfo{person}{Rewon Child}, \bibinfo{person}{Aditya Ramesh}, \bibinfo{person}{Daniel~M. Ziegler}, \bibinfo{person}{Jeffrey Wu}, \bibinfo{person}{Clemens Winter}, \bibinfo{person}{Christopher Hesse}, \bibinfo{person}{Mark Chen}, \bibinfo{person}{Eric Sigler}, \bibinfo{person}{Mateusz Litwin}, \bibinfo{person}{Scott Gray}, \bibinfo{person}{Benjamin Chess}, \bibinfo{person}{Jack Clark}, \bibinfo{person}{Christopher Berner}, \bibinfo{person}{Sam McCandlish}, \bibinfo{person}{Alec Radford}, \bibinfo{person}{Ilya Sutskever},
  {and} \bibinfo{person}{Dario Amodei}.} \bibinfo{year}{2020}\natexlab{}.
\newblock \showarticletitle{Language {Models} are {Few}-{Shot} {Learners}}.
\newblock \bibinfo{journal}{\emph{arXiv:2005.14165 [cs]}} (\bibinfo{date}{June} \bibinfo{year}{2020}).
\newblock
\urldef\tempurl%
\url{http://arxiv.org/abs/2005.14165}
\showURL{%
\tempurl}
\newblock
\shownote{00030 arXiv: 2005.14165.}


\bibitem[\protect\citeauthoryear{Callahan, Freire, Santos, Scheidegger, Silva, and Vo}{Callahan et~al\mbox{.}}{2006}]%
        {callahanVisTrailsVisualizationMeets2006}
\bibfield{author}{\bibinfo{person}{Steven~P. Callahan}, \bibinfo{person}{Juliana Freire}, \bibinfo{person}{Emanuele Santos}, \bibinfo{person}{Carlos~E. Scheidegger}, \bibinfo{person}{Cláudio~T. Silva}, {and} \bibinfo{person}{Huy~T. Vo}.} \bibinfo{year}{2006}\natexlab{}.
\newblock \showarticletitle{{VisTrails}: visualization meets data management}. In \bibinfo{booktitle}{\emph{Proceedings of the 2006 {ACM} {SIGMOD} international conference on {Management} of data}}. \bibinfo{publisher}{ACM}, \bibinfo{address}{Chicago IL USA}, \bibinfo{pages}{745--747}.
\newblock
\showISBNx{978-1-59593-434-5}
\urldef\tempurl%
\url{https://doi.org/10.1145/1142473.1142574}
\showDOI{\tempurl}


\bibitem[\protect\citeauthoryear{Chattopadhyay, Prasad, Henley, Sarma, and Barik}{Chattopadhyay et~al\mbox{.}}{2020}]%
        {chattopadhyayWhatWrongComputational2020b}
\bibfield{author}{\bibinfo{person}{Souti Chattopadhyay}, \bibinfo{person}{Ishita Prasad}, \bibinfo{person}{Austin~Z. Henley}, \bibinfo{person}{Anita Sarma}, {and} \bibinfo{person}{Titus Barik}.} \bibinfo{year}{2020}\natexlab{}.
\newblock \showarticletitle{What's {Wrong} with {Computational} {Notebooks}? {Pain} {Points}, {Needs}, and {Design} {Opportunities}}. In \bibinfo{booktitle}{\emph{Proceedings of the 2020 {CHI} {Conference} on {Human} {Factors} in {Computing} {Systems}}}. \bibinfo{publisher}{ACM}, \bibinfo{address}{Honolulu HI USA}, \bibinfo{pages}{1--12}.
\newblock
\showISBNx{978-1-4503-6708-0}
\urldef\tempurl%
\url{https://doi.org/10.1145/3313831.3376729}
\showDOI{\tempurl}


\bibitem[\protect\citeauthoryear{Chen, Hoffswell, Guo, Rossi, Chan, Du, Koh, and Liu}{Chen et~al\mbox{.}}{[n.d.]}]%
        {chenWHATSNEXTGuidanceenrichedExploratory}
\bibfield{author}{\bibinfo{person}{Chen Chen}, \bibinfo{person}{Jane Hoffswell}, \bibinfo{person}{Shunan Guo}, \bibinfo{person}{Ryan Rossi}, \bibinfo{person}{Yeuk-Yin Chan}, \bibinfo{person}{Fan Du}, \bibinfo{person}{Eunyee Koh}, {and} \bibinfo{person}{Zhicheng Liu}.} \bibinfo{year}{[n.d.]}\natexlab{}.
\newblock \showarticletitle{{WHATSNEXT}: {Guidance}-enriched {Exploratory} {Data} {Analysis} with {Interactive}, {Low}-{Code} {Notebooks}}.
\newblock  (\bibinfo{year}{[n.\,d.]}).
\newblock


\bibitem[\protect\citeauthoryear{Chen, Tworek, Jun, Yuan, Pinto, Kaplan, Edwards, Burda, Joseph, Brockman, Ray, Puri, Krueger, Petrov, Khlaaf, Sastry, Mishkin, Chan, Gray, Ryder, Pavlov, Power, Kaiser, Bavarian, Winter, Tillet, Such, Cummings, Plappert, Chantzis, Barnes, Herbert-Voss, Guss, Nichol, Paino, Tezak, Tang, Babuschkin, Balaji, Jain, Saunders, Hesse, Carr, Leike, Achiam, Misra, Morikawa, Radford, Knight, Brundage, Murati, Mayer, Welinder, McGrew, Amodei, McCandlish, Sutskever, and Zaremba}{Chen et~al\mbox{.}}{2021}]%
        {chenEvaluatingLargeLanguage2021}
\bibfield{author}{\bibinfo{person}{Mark Chen}, \bibinfo{person}{Jerry Tworek}, \bibinfo{person}{Heewoo Jun}, \bibinfo{person}{Qiming Yuan}, \bibinfo{person}{Henrique Ponde de~Oliveira Pinto}, \bibinfo{person}{Jared Kaplan}, \bibinfo{person}{Harri Edwards}, \bibinfo{person}{Yuri Burda}, \bibinfo{person}{Nicholas Joseph}, \bibinfo{person}{Greg Brockman}, \bibinfo{person}{Alex Ray}, \bibinfo{person}{Raul Puri}, \bibinfo{person}{Gretchen Krueger}, \bibinfo{person}{Michael Petrov}, \bibinfo{person}{Heidy Khlaaf}, \bibinfo{person}{Girish Sastry}, \bibinfo{person}{Pamela Mishkin}, \bibinfo{person}{Brooke Chan}, \bibinfo{person}{Scott Gray}, \bibinfo{person}{Nick Ryder}, \bibinfo{person}{Mikhail Pavlov}, \bibinfo{person}{Alethea Power}, \bibinfo{person}{Lukasz Kaiser}, \bibinfo{person}{Mohammad Bavarian}, \bibinfo{person}{Clemens Winter}, \bibinfo{person}{Philippe Tillet}, \bibinfo{person}{Felipe~Petroski Such}, \bibinfo{person}{Dave Cummings}, \bibinfo{person}{Matthias Plappert}, \bibinfo{person}{Fotios Chantzis},
  \bibinfo{person}{Elizabeth Barnes}, \bibinfo{person}{Ariel Herbert-Voss}, \bibinfo{person}{William~Hebgen Guss}, \bibinfo{person}{Alex Nichol}, \bibinfo{person}{Alex Paino}, \bibinfo{person}{Nikolas Tezak}, \bibinfo{person}{Jie Tang}, \bibinfo{person}{Igor Babuschkin}, \bibinfo{person}{Suchir Balaji}, \bibinfo{person}{Shantanu Jain}, \bibinfo{person}{William Saunders}, \bibinfo{person}{Christopher Hesse}, \bibinfo{person}{Andrew~N. Carr}, \bibinfo{person}{Jan Leike}, \bibinfo{person}{Josh Achiam}, \bibinfo{person}{Vedant Misra}, \bibinfo{person}{Evan Morikawa}, \bibinfo{person}{Alec Radford}, \bibinfo{person}{Matthew Knight}, \bibinfo{person}{Miles Brundage}, \bibinfo{person}{Mira Murati}, \bibinfo{person}{Katie Mayer}, \bibinfo{person}{Peter Welinder}, \bibinfo{person}{Bob McGrew}, \bibinfo{person}{Dario Amodei}, \bibinfo{person}{Sam McCandlish}, \bibinfo{person}{Ilya Sutskever}, {and} \bibinfo{person}{Wojciech Zaremba}.} \bibinfo{year}{2021}\natexlab{}.
\newblock \bibinfo{title}{Evaluating {Large} {Language} {Models} {Trained} on {Code}}.
\newblock
\newblock
\urldef\tempurl%
\url{http://arxiv.org/abs/2107.03374}
\showURL{%
\tempurl}
\newblock
\shownote{arXiv:2107.03374 [cs].}


\bibitem[\protect\citeauthoryear{Dow, Glassco, Kass, Schwarz, and Klemmer}{Dow et~al\mbox{.}}{2009}]%
        {dowEffectParallelPrototyping2009}
\bibfield{author}{\bibinfo{person}{S.~P. Dow}, \bibinfo{person}{A. Glassco}, \bibinfo{person}{J. Kass}, \bibinfo{person}{M. Schwarz}, {and} \bibinfo{person}{S.~R. Klemmer}.} \bibinfo{year}{2009}\natexlab{}.
\newblock \bibinfo{booktitle}{\emph{The {Effect} of {Parallel} {Prototyping} on {Design} {Performance}, {Learning}, and {Self}-{Efficacy}}}.
\newblock \bibinfo{type}{{T}echnical {R}eport} September. \bibinfo{institution}{Stanford University}, \bibinfo{address}{Palo Alto, CA}.
\newblock


\bibitem[\protect\citeauthoryear{Dow, Glassco, Kass, Schwarz, Schwartz, and Klemmer}{Dow et~al\mbox{.}}{2010}]%
        {dowParallelPrototypingLeads2010}
\bibfield{author}{\bibinfo{person}{Steven~P. Dow}, \bibinfo{person}{Alana Glassco}, \bibinfo{person}{Jonathan Kass}, \bibinfo{person}{Melissa Schwarz}, \bibinfo{person}{Daniel~L. Schwartz}, {and} \bibinfo{person}{Scott~R. Klemmer}.} \bibinfo{year}{2010}\natexlab{}.
\newblock \showarticletitle{Parallel {Prototyping} {Leads} to {Better} {Design} {Results}, {More} {Divergence}, and {Increased} {Self}-efficacy}.
\newblock \bibinfo{journal}{\emph{ACM Trans. Comput.-Hum. Interact.}} \bibinfo{volume}{17}, \bibinfo{number}{4} (\bibinfo{date}{Dec.} \bibinfo{year}{2010}), \bibinfo{pages}{18:1--18:24}.
\newblock
\showISSN{1073-0516}
\urldef\tempurl%
\url{https://doi.org/10.1145/1879831.1879836}
\showDOI{\tempurl}


\bibitem[\protect\citeauthoryear{Dow, Heddleston, and Klemmer}{Dow et~al\mbox{.}}{[n.d.]}]%
        {dowEfficacyPrototypingTime}
\bibfield{author}{\bibinfo{person}{Steven~P Dow}, \bibinfo{person}{Kate Heddleston}, {and} \bibinfo{person}{Scott~R Klemmer}.} \bibinfo{year}{[n.d.]}\natexlab{}.
\newblock \showarticletitle{The efficacy of prototyping under time constraints}.
\newblock  (\bibinfo{year}{[n.\,d.]}).
\newblock


\bibitem[\protect\citeauthoryear{Girolami, Mischak, and Krebs}{Girolami et~al\mbox{.}}{2006}]%
        {girolamiAnalysisComplexMultidimensional2006}
\bibfield{author}{\bibinfo{person}{Mark Girolami}, \bibinfo{person}{Harald Mischak}, {and} \bibinfo{person}{Ronald Krebs}.} \bibinfo{year}{2006}\natexlab{}.
\newblock \showarticletitle{Analysis of complex, multidimensional datasets}.
\newblock \bibinfo{journal}{\emph{Drug Discovery Today: Technologies}} \bibinfo{volume}{3}, \bibinfo{number}{1} (\bibinfo{date}{March} \bibinfo{year}{2006}), \bibinfo{pages}{13--19}.
\newblock
\showISSN{17406749}
\urldef\tempurl%
\url{https://doi.org/10.1016/j.ddtec.2006.03.010}
\showDOI{\tempurl}


\bibitem[\protect\citeauthoryear{Grammel, Tory, and Storey}{Grammel et~al\mbox{.}}{2010}]%
        {grammelHowInformationVisualization2010}
\bibfield{author}{\bibinfo{person}{L Grammel}, \bibinfo{person}{M Tory}, {and} \bibinfo{person}{M Storey}.} \bibinfo{year}{2010}\natexlab{}.
\newblock \showarticletitle{How {Information} {Visualization} {Novices} {Construct} {Visualizations}}.
\newblock \bibinfo{journal}{\emph{IEEE Transactions on Visualization and Computer Graphics}} \bibinfo{volume}{16}, \bibinfo{number}{6} (\bibinfo{date}{Nov.} \bibinfo{year}{2010}), \bibinfo{pages}{943--952}.
\newblock
\showISSN{1077-2626}
\urldef\tempurl%
\url{https://doi.org/10.1109/TVCG.2010.164}
\showDOI{\tempurl}


\bibitem[\protect\citeauthoryear{Gu, Jun, and Althoff}{Gu et~al\mbox{.}}{2023}]%
        {guUnderstandingSupportingDebugging2023}
\bibfield{author}{\bibinfo{person}{Ken Gu}, \bibinfo{person}{Eunice Jun}, {and} \bibinfo{person}{Tim Althoff}.} \bibinfo{year}{2023}\natexlab{}.
\newblock \bibinfo{title}{Understanding and {Supporting} {Debugging} {Workflows} in {Multiverse} {Analysis}}.
\newblock
\newblock
\urldef\tempurl%
\url{http://arxiv.org/abs/2210.03804}
\showURL{%
\tempurl}
\newblock
\shownote{arXiv:2210.03804 [cs].}


\bibitem[\protect\citeauthoryear{Guo, Karavani, Endert, and Kwon}{Guo et~al\mbox{.}}{2023}]%
        {guoCausalvisVisualizationsCausal2023a}
\bibfield{author}{\bibinfo{person}{Grace Guo}, \bibinfo{person}{Ehud Karavani}, \bibinfo{person}{Alex Endert}, {and} \bibinfo{person}{Bum~Chul Kwon}.} \bibinfo{year}{2023}\natexlab{}.
\newblock \showarticletitle{Causalvis: {Visualizations} for {Causal} {Inference}}. In \bibinfo{booktitle}{\emph{Proceedings of the 2023 {CHI} {Conference} on {Human} {Factors} in {Computing} {Systems}}}. \bibinfo{publisher}{ACM}, \bibinfo{address}{Hamburg Germany}, \bibinfo{pages}{1--20}.
\newblock
\showISBNx{978-1-4503-9421-5}
\urldef\tempurl%
\url{https://doi.org/10.1145/3544548.3581236}
\showDOI{\tempurl}


\bibitem[\protect\citeauthoryear{Horak, Badam, Elmqvist, and Dachselt}{Horak et~al\mbox{.}}{2018}]%
        {horakWhenDavidMeets2018}
\bibfield{author}{\bibinfo{person}{Tom Horak}, \bibinfo{person}{Sriram~Karthik Badam}, \bibinfo{person}{Niklas Elmqvist}, {and} \bibinfo{person}{Raimund Dachselt}.} \bibinfo{year}{2018}\natexlab{}.
\newblock \showarticletitle{When {David} {Meets} {Goliath}: {Combining} {Smartwatches} with a {Large} {Vertical} {Display} for {Visual} {Data} {Exploration}}. In \bibinfo{booktitle}{\emph{Proceedings of the 2018 {CHI} {Conference} on {Human} {Factors} in {Computing} {Systems}}}. \bibinfo{publisher}{ACM}, \bibinfo{address}{Montreal QC Canada}, \bibinfo{pages}{1--13}.
\newblock
\showISBNx{978-1-4503-5620-6}
\urldef\tempurl%
\url{https://doi.org/10.1145/3173574.3173593}
\showDOI{\tempurl}


\bibitem[\protect\citeauthoryear{Hutchins, Hollan, and Norman}{Hutchins et~al\mbox{.}}{[n.d.]}]%
        {hutchinsDirectManipulationInterfaces}
\bibfield{author}{\bibinfo{person}{Edwin~L Hutchins}, \bibinfo{person}{James~D Hollan}, {and} \bibinfo{person}{Donald~A Norman}.} \bibinfo{year}{[n.d.]}\natexlab{}.
\newblock \showarticletitle{Direct {Manipulation} {Interfaces}}.
\newblock  (\bibinfo{year}{[n.\,d.]}).
\newblock


\bibitem[\protect\citeauthoryear{Keim, Mansmann, Schneidewind, and Ziegler}{Keim et~al\mbox{.}}{2006}]%
        {keimChallengesVisualData2006}
\bibfield{author}{\bibinfo{person}{D.A. Keim}, \bibinfo{person}{F. Mansmann}, \bibinfo{person}{J. Schneidewind}, {and} \bibinfo{person}{H. Ziegler}.} \bibinfo{year}{2006}\natexlab{}.
\newblock \showarticletitle{Challenges in {Visual} {Data} {Analysis}}. In \bibinfo{booktitle}{\emph{Tenth {International} {Conference} on {Information} {Visualisation} ({IV}'06)}}. \bibinfo{publisher}{IEEE}, \bibinfo{address}{London, England}, \bibinfo{pages}{9--16}.
\newblock
\showISBNx{978-0-7695-2602-7}
\urldef\tempurl%
\url{https://doi.org/10.1109/IV.2006.31}
\showDOI{\tempurl}


\bibitem[\protect\citeauthoryear{Kerne, Lupfer, Linder, Qu, Valdez, Jain, Keith, Carrasco, Vanegas, and Billingsley}{Kerne et~al\mbox{.}}{2017}]%
        {kerneStrategiesFreeFormWeb2017}
\bibfield{author}{\bibinfo{person}{Andruid Kerne}, \bibinfo{person}{Nic Lupfer}, \bibinfo{person}{Rhema Linder}, \bibinfo{person}{Yin Qu}, \bibinfo{person}{Alyssa Valdez}, \bibinfo{person}{Ajit Jain}, \bibinfo{person}{Kade Keith}, \bibinfo{person}{Matthew Carrasco}, \bibinfo{person}{Jorge Vanegas}, {and} \bibinfo{person}{Andrew Billingsley}.} \bibinfo{year}{2017}\natexlab{}.
\newblock \showarticletitle{Strategies of {Free}-{Form} {Web} {Curation}: {Processes} of {Creative} {Engagement} with {Prior} {Work}}. In \bibinfo{booktitle}{\emph{Proceedings of the 2017 {ACM} {SIGCHI} {Conference} on {Creativity} and {Cognition}}} \emph{(\bibinfo{series}{C\&{C} '17})}. \bibinfo{publisher}{ACM}, \bibinfo{address}{New York, NY, USA}, \bibinfo{pages}{380--392}.
\newblock
\showISBNx{978-1-4503-4403-6}
\urldef\tempurl%
\url{https://doi.org/10.1145/3059454.3059471}
\showDOI{\tempurl}


\bibitem[\protect\citeauthoryear{Kery, Radensky, Arya, John, and Myers}{Kery et~al\mbox{.}}{2018}]%
        {keryStoryNotebookExploratory2018}
\bibfield{author}{\bibinfo{person}{Mary~Beth Kery}, \bibinfo{person}{Marissa Radensky}, \bibinfo{person}{Mahima Arya}, \bibinfo{person}{Bonnie~E. John}, {and} \bibinfo{person}{Brad~A. Myers}.} \bibinfo{year}{2018}\natexlab{}.
\newblock \showarticletitle{The {Story} in the {Notebook}: {Exploratory} {Data} {Science} using a {Literate} {Programming} {Tool}}. In \bibinfo{booktitle}{\emph{Proceedings of the 2018 {CHI} {Conference} on {Human} {Factors} in {Computing} {Systems}}}. \bibinfo{publisher}{ACM}, \bibinfo{address}{Montreal QC Canada}, \bibinfo{pages}{1--11}.
\newblock
\showISBNx{978-1-4503-5620-6}
\urldef\tempurl%
\url{https://doi.org/10.1145/3173574.3173748}
\showDOI{\tempurl}


\bibitem[\protect\citeauthoryear{Knudsen, Jakobsen, and Hornbæk}{Knudsen et~al\mbox{.}}{2012}]%
        {knudsenExploratoryStudyHow2012}
\bibfield{author}{\bibinfo{person}{Søren Knudsen}, \bibinfo{person}{Mikkel~Rønne Jakobsen}, {and} \bibinfo{person}{Kasper Hornbæk}.} \bibinfo{year}{2012}\natexlab{}.
\newblock \showarticletitle{An exploratory study of how abundant display space may support data analysis}. In \bibinfo{booktitle}{\emph{Proceedings of the 7th {Nordic} {Conference} on {Human}-{Computer} {Interaction}: {Making} {Sense} {Through} {Design}}}. \bibinfo{publisher}{ACM}, \bibinfo{address}{Copenhagen Denmark}, \bibinfo{pages}{558--567}.
\newblock
\showISBNx{978-1-4503-1482-4}
\urldef\tempurl%
\url{https://doi.org/10.1145/2399016.2399102}
\showDOI{\tempurl}


\bibitem[\protect\citeauthoryear{Liu, Kale, Althoff, and Heer}{Liu et~al\mbox{.}}{2021}]%
        {liuBobaAuthoringVisualizing2021}
\bibfield{author}{\bibinfo{person}{Yang Liu}, \bibinfo{person}{Alex Kale}, \bibinfo{person}{Tim Althoff}, {and} \bibinfo{person}{Jeffrey Heer}.} \bibinfo{year}{2021}\natexlab{}.
\newblock \showarticletitle{Boba: {Authoring} and {Visualizing} {Multiverse} {Analyses}}.
\newblock \bibinfo{journal}{\emph{IEEE Transactions on Visualization and Computer Graphics}} \bibinfo{volume}{27}, \bibinfo{number}{2} (\bibinfo{date}{Feb.} \bibinfo{year}{2021}), \bibinfo{pages}{1753--1763}.
\newblock
\showISSN{1077-2626, 1941-0506, 2160-9306}
\urldef\tempurl%
\url{https://doi.org/10.1109/TVCG.2020.3028985}
\showDOI{\tempurl}


\bibitem[\protect\citeauthoryear{Lupfer, Kerne, Webb, and Linder}{Lupfer et~al\mbox{.}}{2016}]%
        {lupferPatternsFreeformCuration2016a}
\bibfield{author}{\bibinfo{person}{Nic Lupfer}, \bibinfo{person}{Andruid Kerne}, \bibinfo{person}{Andrew~M. Webb}, {and} \bibinfo{person}{Rhema Linder}.} \bibinfo{year}{2016}\natexlab{}.
\newblock \showarticletitle{Patterns of {Free}-form {Curation}: {Visual} {Thinking} with {Web} {Content}}. In \bibinfo{booktitle}{\emph{Proceedings of the 24th {ACM} international conference on {Multimedia}}}. \bibinfo{publisher}{ACM}, \bibinfo{address}{Amsterdam The Netherlands}, \bibinfo{pages}{12--21}.
\newblock
\showISBNx{978-1-4503-3603-1}
\urldef\tempurl%
\url{https://doi.org/10.1145/2964284.2964303}
\showDOI{\tempurl}
\newblock
\shownote{00030.}


\bibitem[\protect\citeauthoryear{MacNeil, Huang, Chen, Ding, Yu, Nakai, and Dow}{MacNeil et~al\mbox{.}}{2023}]%
        {macneilFreeformTemplatesCombining2023}
\bibfield{author}{\bibinfo{person}{Stephen MacNeil}, \bibinfo{person}{Ziheng Huang}, \bibinfo{person}{Kenneth Chen}, \bibinfo{person}{Zijian Ding}, \bibinfo{person}{Alex Yu}, \bibinfo{person}{Kendall Nakai}, {and} \bibinfo{person}{Steven~P. Dow}.} \bibinfo{year}{2023}\natexlab{}.
\newblock \showarticletitle{Freeform {Templates}: {Combining} {Freeform} {Curation} with {Structured} {Templates}}. In \bibinfo{booktitle}{\emph{Creativity and {Cognition}}}. \bibinfo{pages}{478--488}.
\newblock
\urldef\tempurl%
\url{https://doi.org/10.1145/3591196.3593337}
\showDOI{\tempurl}
\newblock
\shownote{arXiv:2305.00937 [cs].}


\bibitem[\protect\citeauthoryear{Neumann, Tang, and Carpendale}{Neumann et~al\mbox{.}}{[n.d.]}]%
        {neumannFrameworkVisualInformation}
\bibfield{author}{\bibinfo{person}{Petra Neumann}, \bibinfo{person}{Anthony Tang}, {and} \bibinfo{person}{Sheelagh Carpendale}.} \bibinfo{year}{[n.d.]}\natexlab{}.
\newblock \showarticletitle{A {Framework} for {Visual} {Information} {Analysis}}.
\newblock  (\bibinfo{year}{[n.\,d.]}).
\newblock


\bibitem[\protect\citeauthoryear{Nielsen}{Nielsen}{1993}]%
        {nielsenIterativeUserinterfaceDesign1993}
\bibfield{author}{\bibinfo{person}{J. Nielsen}.} \bibinfo{year}{1993}\natexlab{}.
\newblock \showarticletitle{Iterative user-interface design}.
\newblock \bibinfo{journal}{\emph{Computer}} \bibinfo{volume}{26}, \bibinfo{number}{11} (\bibinfo{date}{Nov.} \bibinfo{year}{1993}), \bibinfo{pages}{32--41}.
\newblock
\showISSN{0018-9162}
\urldef\tempurl%
\url{https://doi.org/10.1109/2.241424}
\showDOI{\tempurl}


\bibitem[\protect\citeauthoryear{Russell, Stefik, Pirolli, and Card}{Russell et~al\mbox{.}}{1993}]%
        {russellCostStructureSensemaking1993}
\bibfield{author}{\bibinfo{person}{Daniel~M. Russell}, \bibinfo{person}{Mark~J. Stefik}, \bibinfo{person}{Peter Pirolli}, {and} \bibinfo{person}{Stuart~K. Card}.} \bibinfo{year}{1993}\natexlab{}.
\newblock \showarticletitle{The {Cost} {Structure} of {Sensemaking}}. In \bibinfo{booktitle}{\emph{Proceedings of the {INTERACT} '93 and {CHI} '93 {Conference} on {Human} {Factors} in {Computing} {Systems}}} \emph{(\bibinfo{series}{{CHI} '93})}. \bibinfo{publisher}{ACM}, \bibinfo{address}{New York, NY, USA}, \bibinfo{pages}{269--276}.
\newblock
\showISBNx{0-89791-575-5}
\urldef\tempurl%
\url{https://doi.org/10.1145/169059.169209}
\showDOI{\tempurl}


\bibitem[\protect\citeauthoryear{Sarma, Kale, Moon, Taback, Chevalier, Hullman, and Kay}{Sarma et~al\mbox{.}}{2021}]%
        {sarmaMultiverseMultiplexingAlternative2021}
\bibfield{author}{\bibinfo{person}{Abhraneel Sarma}, \bibinfo{person}{Alex Kale}, \bibinfo{person}{Michael~Jongho Moon}, \bibinfo{person}{Nathan Taback}, \bibinfo{person}{Fanny Chevalier}, \bibinfo{person}{Jessica Hullman}, {and} \bibinfo{person}{Matthew Kay}.} \bibinfo{year}{2021}\natexlab{}.
\newblock \bibinfo{booktitle}{\emph{multiverse: {Multiplexing} {Alternative} {Data} {Analyses} in {R} {Notebooks}}}.
\newblock \bibinfo{type}{preprint}. \bibinfo{institution}{Open Science Framework}.
\newblock
\urldef\tempurl%
\url{https://doi.org/10.31219/osf.io/yfbwm}
\showDOI{\tempurl}


\bibitem[\protect\citeauthoryear{Satyanarayan and Heer}{Satyanarayan and Heer}{2014}]%
        {satyanarayanLyraInteractiveVisualization2014}
\bibfield{author}{\bibinfo{person}{Arvind Satyanarayan} {and} \bibinfo{person}{Jeffrey Heer}.} \bibinfo{year}{2014}\natexlab{}.
\newblock \showarticletitle{Lyra: {An} {Interactive} {Visualization} {Design} {Environment}}.
\newblock \bibinfo{journal}{\emph{Computer Graphics Forum}} \bibinfo{volume}{33}, \bibinfo{number}{3} (\bibinfo{date}{June} \bibinfo{year}{2014}), \bibinfo{pages}{351--360}.
\newblock
\showISSN{0167-7055, 1467-8659}
\urldef\tempurl%
\url{https://doi.org/10.1111/cgf.12391}
\showDOI{\tempurl}


\bibitem[\protect\citeauthoryear{Silva, Freire, and Callahan}{Silva et~al\mbox{.}}{2007}]%
        {silva2007provenance}
\bibfield{author}{\bibinfo{person}{Claudio~T. Silva}, \bibinfo{person}{Juliana Freire}, {and} \bibinfo{person}{Steven~P. Callahan}.} \bibinfo{year}{2007}\natexlab{}.
\newblock \showarticletitle{Provenance for {Visualizations}: {Reproducibility} and {Beyond}}.
\newblock \bibinfo{journal}{\emph{Computing in Science \& Engineering}} \bibinfo{volume}{9}, \bibinfo{number}{5} (\bibinfo{date}{Sept.} \bibinfo{year}{2007}), \bibinfo{pages}{82--89}.
\newblock
\showISSN{1521-9615}
\urldef\tempurl%
\url{https://doi.org/10.1109/MCSE.2007.106}
\showDOI{\tempurl}


\bibitem[\protect\citeauthoryear{{Texas Tech University}, Baker, Jones, {Texas Tech University}, Burkman, and {Oklahoma State University}}{{Texas Tech University} et~al\mbox{.}}{2009}]%
        {texastechuniversityUsingVisualRepresentations2009}
\bibfield{author}{\bibinfo{person}{{Texas Tech University}}, \bibinfo{person}{Jeff Baker}, \bibinfo{person}{Donald Jones}, \bibinfo{person}{{Texas Tech University}}, \bibinfo{person}{Jim Burkman}, {and} \bibinfo{person}{{Oklahoma State University}}.} \bibinfo{year}{2009}\natexlab{}.
\newblock \showarticletitle{Using {Visual} {Representations} of {Data} to {Enhance} {Sensemaking} in {Data} {Exploration} {Tasks}}.
\newblock \bibinfo{journal}{\emph{Journal of the Association for Information Systems}} \bibinfo{volume}{10}, \bibinfo{number}{7} (\bibinfo{date}{July} \bibinfo{year}{2009}), \bibinfo{pages}{533--559}.
\newblock
\showISSN{15369323}
\urldef\tempurl%
\url{https://doi.org/10.17705/1jais.00204}
\showDOI{\tempurl}


\bibitem[\protect\citeauthoryear{Vuillemot and Boy}{Vuillemot and Boy}{2018}]%
        {vuillemotStructuringVisualizationMockUps2018}
\bibfield{author}{\bibinfo{person}{Romain Vuillemot} {and} \bibinfo{person}{Jeremy Boy}.} \bibinfo{year}{2018}\natexlab{}.
\newblock \showarticletitle{Structuring {Visualization} {Mock}-{Ups} at the {Graphical} {Level} by {Dividing} the {Display} {Space}}.
\newblock \bibinfo{journal}{\emph{IEEE Transactions on Visualization and Computer Graphics}} \bibinfo{volume}{24}, \bibinfo{number}{1} (\bibinfo{date}{Jan.} \bibinfo{year}{2018}), \bibinfo{pages}{424--434}.
\newblock
\showISSN{1077-2626}
\urldef\tempurl%
\url{https://doi.org/10.1109/TVCG.2017.2743998}
\showDOI{\tempurl}


\bibitem[\protect\citeauthoryear{Wang, Dai, and Edwards}{Wang et~al\mbox{.}}{2022}]%
        {wang2022stickyland}
\bibfield{author}{\bibinfo{person}{Zijie~J Wang}, \bibinfo{person}{Katie Dai}, {and} \bibinfo{person}{W~Keith Edwards}.} \bibinfo{year}{2022}\natexlab{}.
\newblock \showarticletitle{Stickyland: Breaking the linear presentation of computational notebooks}. In \bibinfo{booktitle}{\emph{CHI Conference on Human Factors in Computing Systems Extended Abstracts}}. \bibinfo{pages}{1--7}.
\newblock


\bibitem[\protect\citeauthoryear{Weinman, Drucker, Barik, and DeLine}{Weinman et~al\mbox{.}}{2021}]%
        {weinmanForkItSupporting2021}
\bibfield{author}{\bibinfo{person}{Nathaniel Weinman}, \bibinfo{person}{Steven~M. Drucker}, \bibinfo{person}{Titus Barik}, {and} \bibinfo{person}{Robert DeLine}.} \bibinfo{year}{2021}\natexlab{}.
\newblock \showarticletitle{Fork {It}: {Supporting} {Stateful} {Alternatives} in {Computational} {Notebooks}}. In \bibinfo{booktitle}{\emph{Proceedings of the 2021 {CHI} {Conference} on {Human} {Factors} in {Computing} {Systems}}}. \bibinfo{publisher}{ACM}, \bibinfo{address}{Yokohama Japan}, \bibinfo{pages}{1--12}.
\newblock
\showISBNx{978-1-4503-8096-6}
\urldef\tempurl%
\url{https://doi.org/10.1145/3411764.3445527}
\showDOI{\tempurl}


\bibitem[\protect\citeauthoryear{Wongsuphasawat, Moritz, Anand, Mackinlay, Howe, and Heer}{Wongsuphasawat et~al\mbox{.}}{2016}]%
        {wongsuphasawatVoyagerExploratoryAnalysis2016}
\bibfield{author}{\bibinfo{person}{Kanit Wongsuphasawat}, \bibinfo{person}{Dominik Moritz}, \bibinfo{person}{Anushka Anand}, \bibinfo{person}{Jock Mackinlay}, \bibinfo{person}{Bill Howe}, {and} \bibinfo{person}{Jeffrey Heer}.} \bibinfo{year}{2016}\natexlab{}.
\newblock \showarticletitle{Voyager: {Exploratory} {Analysis} via {Faceted} {Browsing} of {Visualization} {Recommendations}}.
\newblock \bibinfo{journal}{\emph{IEEE Transactions on Visualization and Computer Graphics}} \bibinfo{volume}{22}, \bibinfo{number}{1} (\bibinfo{date}{Jan.} \bibinfo{year}{2016}), \bibinfo{pages}{649--658}.
\newblock
\showISSN{1077-2626}
\urldef\tempurl%
\url{https://doi.org/10.1109/TVCG.2015.2467191}
\showDOI{\tempurl}


\bibitem[\protect\citeauthoryear{Wongsuphasawat, Qu, Moritz, Chang, Ouk, Anand, Mackinlay, Howe, and Heer}{Wongsuphasawat et~al\mbox{.}}{2017}]%
        {wongsuphasawatVoyagerAugmentingVisual2017}
\bibfield{author}{\bibinfo{person}{Kanit Wongsuphasawat}, \bibinfo{person}{Zening Qu}, \bibinfo{person}{Dominik Moritz}, \bibinfo{person}{Riley Chang}, \bibinfo{person}{Felix Ouk}, \bibinfo{person}{Anushka Anand}, \bibinfo{person}{Jock Mackinlay}, \bibinfo{person}{Bill Howe}, {and} \bibinfo{person}{Jeffrey Heer}.} \bibinfo{year}{2017}\natexlab{}.
\newblock \showarticletitle{Voyager 2: {Augmenting} {Visual} {Analysis} with {Partial} {View} {Specifications}}. In \bibinfo{booktitle}{\emph{Proceedings of the 2017 {CHI} {Conference} on {Human} {Factors} in {Computing} {Systems}}}. \bibinfo{publisher}{ACM}, \bibinfo{address}{Denver Colorado USA}, \bibinfo{pages}{2648--2659}.
\newblock
\showISBNx{978-1-4503-4655-9}
\urldef\tempurl%
\url{https://doi.org/10.1145/3025453.3025768}
\showDOI{\tempurl}


\bibitem[\protect\citeauthoryear{Wu, Hellerstein, and Satyanarayan}{Wu et~al\mbox{.}}{2020}]%
        {wuB2BridgingCode2020}
\bibfield{author}{\bibinfo{person}{Yifan Wu}, \bibinfo{person}{Joseph~M. Hellerstein}, {and} \bibinfo{person}{Arvind Satyanarayan}.} \bibinfo{year}{2020}\natexlab{}.
\newblock \showarticletitle{B2: {Bridging} {Code} and {Interactive} {Visualization} in {Computational} {Notebooks}}. In \bibinfo{booktitle}{\emph{Proceedings of the 33rd {Annual} {ACM} {Symposium} on {User} {Interface} {Software} and {Technology}}}. \bibinfo{publisher}{ACM}, \bibinfo{address}{Virtual Event USA}, \bibinfo{pages}{152--165}.
\newblock
\showISBNx{978-1-4503-7514-6}
\urldef\tempurl%
\url{https://doi.org/10.1145/3379337.3415851}
\showDOI{\tempurl}


\bibitem[\protect\citeauthoryear{Zong, Barnwal, Neogy, and Satyanarayan}{Zong et~al\mbox{.}}{2020}]%
        {zongLyraDesigningInteractive2020}
\bibfield{author}{\bibinfo{person}{Jonathan Zong}, \bibinfo{person}{Dhiraj Barnwal}, \bibinfo{person}{Rupayan Neogy}, {and} \bibinfo{person}{Arvind Satyanarayan}.} \bibinfo{year}{2020}\natexlab{}.
\newblock \bibinfo{title}{Lyra 2: {Designing} {Interactive} {Visualizations} by {Demonstration}}.
\newblock
\newblock
\urldef\tempurl%
\url{http://arxiv.org/abs/2008.09576}
\showURL{%
\tempurl}
\newblock
\shownote{arXiv:2008.09576 [cs].}


\end{thebibliography}
